\documentclass{article}
\usepackage[utf8]{inputenc}
\usepackage{amsmath}
\usepackage{graphicx}

\usepackage{amssymb,amsthm,amsmath}

\title{Utilising a learned forward operator in the inverse problem of photoacoustic tomography}
\author{Karoliina Puronhaara$^{1}$,
Teemu Sahlström$^{1}$, \\
Andreas Hauptmann$^{2,3}$ and 
Tanja Tarvainen${^1}$}

\date{\small \textit{$^1$Department of Technical Physics, University of Eastern Finland, Finland \\ \medskip
$^2$Research Unit of Mathematical Sciences, University of Oulu, Finland \\ \medskip
$^3$Department of Computer Science, University College London, United Kingdom}}

\begin{document}
\maketitle 
\begin{abstract}
We study the use of a learned forward operator in the inverse problem of photoacoustic tomography. The Fourier neural operator to approximate the photoacoustic wave propagation is used. Further, the inverse problem is solved using a gradient-based approach with automatic differentiation. The methodology is evaluated using numerical simulations, and the results are compared to a conventional approach, where the forward operator is approximated using the pseudospectral $k$-space method. The results show that the learned forward operator can be used to approximate the photoacoustic wave propagation with good accuracy, and that it can be utilised as a computationally efficient forward operator in solving the inverse problem of photoacoustic tomography.
\end{abstract} 

\section{Introduction}
Photoacoustic tomography (PAT) is an imaging modality based on coupled physics of light and ultrasound \cite{Beard2011, Poudel2019, Wang2016}. In PAT, a short light pulse is directed to the imaged target, generating localised increases in pressure via the photoacoustic effect. This pressure increase relaxes as propagating ultrasound waves that are measured on the boundary of the target. In the inverse problem of PAT, the pressure distribution, referred to as the initial pressure, is estimated from the measured photoacoustic data. 

The inverse problem of PAT has been approached using various methods such as back-projection \cite{Cox2007, Finch2004, Haltmeier2005, Kunyansky2007, Xu2005, Xu2003}, time-reversal \cite{Burgholzer2007,  Hristova2008, Treeby2010, Xu2004}, regularised least-squares techniques \cite{Arridge2016, Dean-Ben2012, Frikel2018, Kultima2026, Wang2013, Wang2012}, and the Bayesian approach \cite{Sahlstrom2020, Tick2018, Tick2016}. Out of these methods, the regularised least squares, Bayesian methods, and time-reversal, utilise a numerical approximation of the model for ultrasound propagation. 

In addition to the classical methods, deep learning has been utilised in PAT in various problems \cite{Grohl2021, Hauptmann2020, Yang2020}. In pre-processing methods, a neural network is used to improve the photoacoustic data before reconstructions \cite{Allman2018}, whereas in post-processing methods, a neural network is used to correct errors in the reconstructed images \cite{Antholzer2019,Guan2020}. Furthermore, in the so-called fully learned methods, a neural network is trained to perform entire image reconstruction process \cite{Waibel2018}. In addition, deep learning has also been utilised in learned iterative reconstructions, where the inverse problem of PAT is solved iteratively with a learned updating operator \cite{Arridge2019,Hauptmann2018}. 

The forward model typically used in PAT is the acoustic wave equation, which is a second-order partial differential equation (PDE). Analytical solutions of the wave equation can be utilised in PAT in specific geometries \cite{Kuchment2008}. Further, numerical approximations, such as, finite difference and pseudospectral $k$-space method are used \cite{Cox2004, Mitsuhashi2017, Poudel2019}. Although, these methods can be implemented efficiently, approximating the forward solution in general cases still remains as one of the computational bottlenecks in PAT. Lately, various deep learning methods for approximating the solutions of the PDEs have been proposed. Some of the most popular approaches include physics informed neural networks (PINN) \cite{Raissi2019} and neural operators (NO) \cite{Lu2021}. PINNs can provide accurate results for various problems by utilising PDE inspired loss functions. However, generally PINNs have to be trained separately for each new instance of parameters. Neural operators, on the other hand, aim to learn the solution operator of a PDE from data and do not utilise the underlying PDE. Furthermore, once trained, the NOs can be used to approximate solutions for large range of parameters. 

Recently, a subclass of the NOs known as the Fourier neural operator (FNO), has been utilised in modelling of ultrasound propagation in PAT \cite{Guan2023}. The FNO-based learned forward operator has further been used as a part of a learned iterative reconstruction approach, where the search direction was approximated using a convolutional neural network and gradients of the objective functional were approximated using a pixel-interpolation approach \cite{Hsu2023}.

In this work, we propose an approach for solving the inverse problem of PAT using a learned forward operator. We approach the inverse problem in the framework of Bayesian inverse problems, and solve the \emph{maximum a posteriori} (MAP) estimate using the Broyden-Fletcher-Goldfarb-Shanno (BFGS) method. The forward model is based on the FNO, and its gradients are computed using automatic differentiation, enabling computationally efficient evaluation of the forward operator and gradients during the optimisation algorithm.

The remainder of this paper is organised as follows. The forward and inverse problems of PAT are reviewed in Section \ref{2}. Then, in Section \ref{3}, a framework for the learned forward operator and an approach utilising that in the inverse problem of PAT are presented. Numerical simulations and their results are presented in Section \ref{4}. Finally, discussion and conclusions of the results are given in Section \ref{5}. 

% The title of section 2:
\section{Photoacoustic tomography}\label{2}
We study the inverse problem of PAT, where the aim is to estimate the initial pressure from photoacoustic data measured from the boundary of the target. The discretised observation model for PAT can be written as
\begin{equation}\label{ObservationModel}
 p_t = Kp_0 + e,   
\end{equation}
where $p_t \in \mathbb{R}^{MH \times 1}$ is a vector of the photoacoustic data measured on M ultrasound sensors at H time instances, $p_0 \in \mathbb{R}^{N \times 1}$ is a unknown initial pressure distribution with N spatial discretisation points, $e \in \mathbb{R}^{MH \times 1}$ is a vector of additive noise, and $K \in \mathbb{R}^{MH \times N}$ is a discretised forward operator. 
In PAT, the forward problem is an acoustic initial value problem, where propagation of ultrasound waves is modelled using the acoustic wave equation. The forward model can be written (in a continuous form) as \cite{Cox2007b}
\begin{equation} \label{eq:waveEquation}
\begin{cases}
      \nabla^2p(r,t)  - \frac{1}{c^2}\frac{\partial^2p(r,t)}{\partial t^2} = 0, \\ 
      p(r, t = 0)  = p_0(r),  \\
      \frac{\partial}{\partial t} p(r,t=0) = 0, \\
\end{cases} 
\end{equation}
where $p(r,t)$ is the pressure,  $r$ is the spatial position, $t$ is the time, $c$ is the speed of sound, and $p_0(r)$ is the initial pressure.

The initial pressure distribution is estimated by computing the \emph{maximum a posteriori} (MAP) estimate \cite{Kaipio2005}
\begin{equation}
    \begin{aligned} \label{eq:MAP1}
    p_{0,\mathrm{MAP}}  & =  \operatorname*{arg \, min}_{p_0} \left\{ \frac{1}{2}\Vert L_e(p_t - Kp_0 -\eta_e)\Vert^2 + \frac{1}{2} \Vert L_{p_0}(p_0 - \eta_{p_0}) \Vert^2\right\} \\  & = \operatorname*{arg \, min}_{p_0} \left\{ f_K(p_0)\right\},
\end{aligned}
\end{equation}
where noise and prior have been modelled as mutually independent and Gaussian distributed, i.e. $e\sim \mathcal{N}(\eta_e, \Gamma_e)$ and $p_0\sim \mathcal{N}(\eta_{p_0},\Gamma_{p_0})$. Further, $L_e$ and $L_{p_0}$ are the Cholesky decompositions of the inverse covariance matrices of the noise and prior distributions such that $L_e^TL_e = \Gamma_e^{-1}$ and $L_{p_0}^TL_{p_0} = \Gamma_{p_0}^{-1}$, and $f_K$ denotes the objective function. 
In this work, we use the BFGS \cite{Nocedal2006} method to solve minimisation problem \eqref{eq:MAP1}. In the BFGS algorithm, the minimisation problem is solved iteratively
\begin{equation}\label{method}
    p_{0, i+1} = p_{0,i} + \alpha_i d_i,
\end{equation}
where $\alpha_i$ is a step length and $d_i$ is a search direction
\begin{equation}
    d_i = -(B_i)^{-1}\nabla f_K(p_{0,i}),
\end{equation}
where the gradient of the objective function $f_K$ is  
\begin{equation}\label{eq:Gradient1}
    \nabla f_K(p_{0,i}) = -K^\textrm{T} \Gamma_e^{-1}(p_t - Kp_{0,i} - \eta_e) + \Gamma_{p_0}^{-1}(p_{0,i} - \eta_{p_0}).
\end{equation}
Further, $B_i$ is an approximation of the Hessian matrix
\begin{equation} \label{eq:B}
    B_{i+1} = B_i - \frac{B_iq_iq_i^\textrm{T}B_i}{q_i^\textrm{T}B_iq_i}+\frac{y_iy_i^\textrm{T}}{y_i^\textrm{T}q_i},
\end{equation}
where
\begin{align}
    q_i &= p_{0,i+1}-p_{0,i},  \label{eq:q}\\
    y_i &= \nabla f_K(p_{0,i+1}) - \nabla f_K(p_{0,i}).
\end{align}

% The title of section 3:
\section{Utilising the Fourier neural operator in photoacoustic tomography} \label{3}
\subsection{Fourier neural operator}
The FNO is a deep learning model designed for modelling solution operators of PDEs \cite{Li2021}. Let us consider, in a general setting, an input function space $a(x) \subset \mathcal{A}$, such as the coefficients or an initial condition of a PDE, and an output or solution function space $u(x) \subset \mathcal{U}$. The aim of the FNO is to learn a mapping $\mathcal{G}: \mathcal{A} \rightarrow \mathcal{U}$ from the input coefficients of a PDE to its solution. The FNO is constructed as an iterative architecture of the form $v_0 \rightarrow v_1 \rightarrow \dots \rightarrow v_J$, where layers $v_j$ can be written as 
\begin{equation}\label{eq:layer}
    v_{j+1}(r) = \sigma(Wv_j(r) + \mathcal{F}^{-1}\{R\mathcal{F}\{v_j\}\}(r)),  \quad \mathrm{where} \quad j = 0, \dots J-1,
\end{equation}
where $\sigma$ is an activation function, and $W$ and $R$ are linear transformations. In addition, $\mathcal{F}$ and $\mathcal{F}^{-1}$ are the Fourier and the inverse Fourier transforms, respectively. In addition to the layers \eqref{eq:layer}, the FNO is equipped with two projection layers, $P$ and $Q$, transforming the inputs and outputs such that $v_0 = P(a(r))$ and $u(r) = Q(v_J(r))$. Furthermore, the mapping $R$ in the Fourier space can be used to truncate the Fourier modes. 

In this work, the FNO is trained to approximate the solution operator of the acoustic wave equation \eqref{eq:waveEquation}. In this case, the input space consists of the initial pressure distribution, and the output is the ultrasound wave-fields in the computational domain. 

\subsection{Inverse problem with a learned forward operator}
Even though the acoustic wave equation \eqref{eq:waveEquation} is linear, neural networks are in general non-linear mappings. The observation model using the learned forward operator is therefore non-linear 
\begin{equation}
    p_t = \Lambda(p_0) + e,
\end{equation}
where $\Lambda(p_0): \mathbb{R}^{N \times 1} \rightarrow \mathbb{R}^{MH \times 1}$ is the learned forward operator. The MAP estimate can then be computed by solving a minimisation problem 
\begin{equation} \label{eq:MAP2}
\begin{aligned} 
    p_{0,\mathrm{MAP}}  &=  \operatorname*{arg \, min}_{p_0} \left\{ \frac{1}{2}\Vert L_e(p_t - \Lambda(p_0) -\eta_e)\Vert^2 + \frac{1}{2} \Vert L_{p_0}(p_0 - \eta_{p_0}) \Vert^2\right\} \\  &= \operatorname*{arg \, min}_{p_0} \left\{ f_{\Lambda}(p_0)\right\},
\end{aligned}
\end{equation}
where $f_{\Lambda}$ denotes the objective function. 

The solution to the minimisation problem \eqref{eq:MAP2} is computed using the BFGS method where the approximation of Hessian matrix $B_i$ is given by \eqref{eq:B}, where
\begin{align}
    q_i &= p_{0,i+1}-p_{0,i}, \\
    y_i &= \nabla f_\Lambda(p_{0,i+1}) - \nabla f_\Lambda(p_{0,i}).
\end{align}
The gradient of the objective function can be expressed in terms of the Jacobian matrix $J_{\Lambda(p_{0,i})}$ of the forward operator 
\begin{equation} \label{eq:Gradient2}
    \nabla f_\Lambda(p_{0,i}) = -J_{\Lambda(p_{0,i})}^\textrm{T} (\Gamma_e^{-1}(p_t - \Lambda(p_{0,i}) - \eta_e)) + \Gamma_{p_0}^{-1}(p_{0,i} - \eta_{p_0}). 
\end{equation}
In this work, we compute the Jacobian vector product in Equation \eqref{eq:Gradient2} using the inbuilt Pytorch function 
\texttt{torch.autograd.grad(x, y, b)}, where \texttt{x = $\Lambda(p_0)$, y = $p_0$} and \texttt{b = $ \Gamma_e^{-1}(p_t - \Lambda(p_{0,i}) - \eta_e)$}.

% The title of section 4:
\section{Simulations} \label{4}
\subsection{Data simulation}
We studied the use of the FNO in the inverse problem of PAT using numerical simulations. In the simulations, a 2D square domain of the size 10 mm $\times$ 10 mm was considered. First, the accuracy of the FNO for simulating propagation of the photoacoustic wave-field in the domain was studied. Then, its use in solving the inverse problem was investigated. Three ultrasound sensor geometries were studied: a full view geometry and two limited view geometries where sensors were modelled on two adjacent sides and on one side of the domain. The numerical phantoms used for training and testing were initial pressure distributions of blood vessel like structures based on the High-Resolution Fundus Image Database \cite{Budai2013, Sahlstrom2023}. Furthermore, generalisation of the methodology was studied using the Shepp-Logan phantom. Throughout the simulations, the speed of sound was $c$ = 1500 m/s. Propagation of the photoacoustic wave-field was simulated using the wave equation \eqref{eq:waveEquation} using the pseudospectral $k$-space method implemented in the k-Wave MATLAB toolbox \cite{Treeby2010Kwave}. Discretisation used in the simulations is described in Table \ref{SimParameters}. Uncorrelated Gaussian zero mean noise with a standard deviation of 1$\%$ of the maximum simulated amplitude was added to the simulated photoacoustic wave-field. After data simulation, the initial pressure distributions and the corresponding simulated photoacoustic wave-fields were linearly interpolated to the discretisation used in solving the inverse problem (Table \ref{SimParameters}). In total, 4000 training, 500 validation and 100 testing pairs of initial pressure distributions and photoacoustic wave-fields were simulated. 

\begin{table}[tbp]
\centering
\caption{Discretisations used in photoacoustic data simulation and solving the inverse problem. Number of pixels $N$, pixel size $\Delta x$, size of the perfectly matching layer (PML, pixels), number of time points $M$ and time step $\Delta t$.
}\label{SimParameters}
\begin{tabular}{llccccccccc}
\hline
                &  & $N$            &  & $\Delta x $ $(\mu$m) &  & PML &  & $M$ &  & $\Delta t$ (ns) \\ \hline
Data simulation      &  & $230^2$ &  & 43.5  &  & 13  &  & 1085  &  & 8.7             \\
Inverse problem &  & $128^2$ &  & 78.1  &  & 8   &  & 256   &  & 37              \\ \hline
\end{tabular}
\end{table}

\subsection{Fourier neural operator architecture}
\begin{figure}[tbp]
    \centering
    \includegraphics[width=1\textwidth]{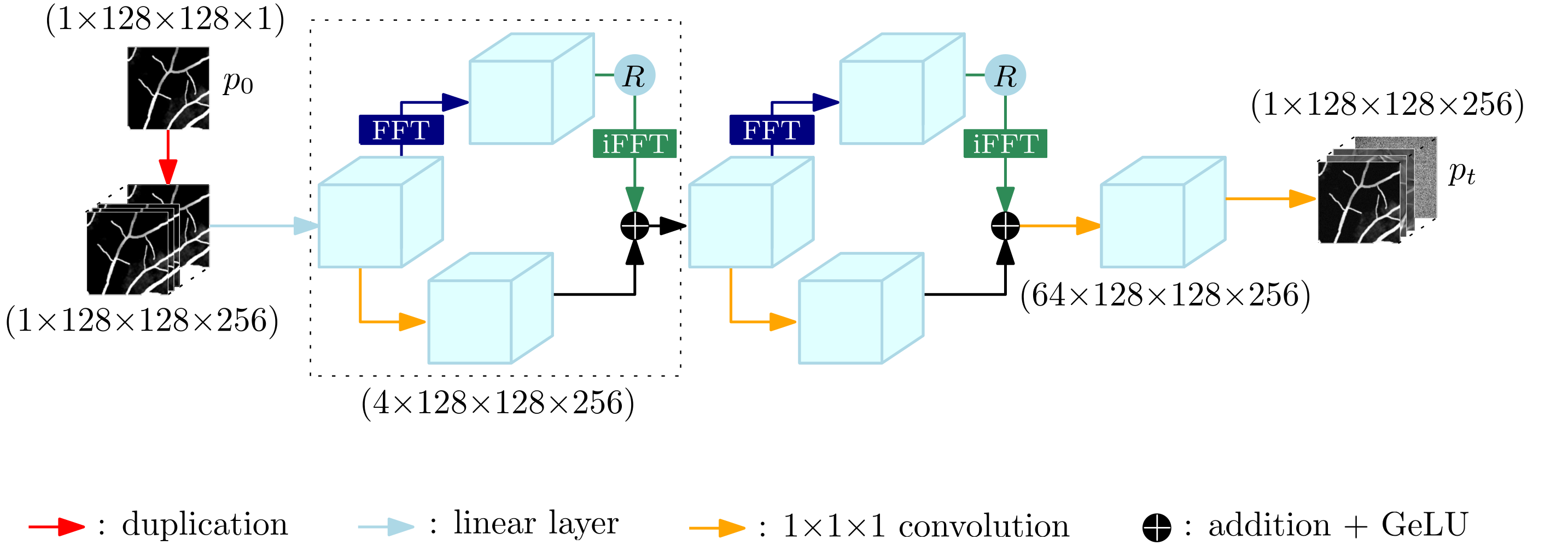}
    \caption{Neural network architecture of the FNO. The numbers in the parentheses indicate the tensor dimensions. One complete Fourier layer is indicated using a dashed box.}
    \label{fig:FNO}
\end{figure}

The FNO used in this work to simulate photoacoustic wave propagation was constructed as a 3D network with two spatial and one temporal dimension. The neural network architecture is illustrated in Figure \ref{fig:FNO}. In the beginning of the network, the initial pressure distribution of size $128 \times 128$ is transformed to the dimensions of the output $128 \times 128 \times 256$ using duplication. The input is then expanded to a higher channel dimension using a linear layer, after which the Fourier layers are applied each with two paths. The first path includes the Fourier transform, followed by a learned linear transformation $R$, that can be used to select a desired number of Fourier modes. At the end of this path, the data is transformed back to the original space using the inverse Fourier transform. In the second path, convolution layers are used. The outputs of both paths are then combined, and a Gaussian Error Linear Unit activation function is applied. After the Fourier layers, two convolution layers are used to map the output of the last Fourier layer to the output of size $128 \times 128 \times 256$. Due to the 3D structure of the network, the entire photoacoustic wave-field in the whole computational domain is obtained using one forward pass.

Important hyperparameters that influence the number of learnable parameters in the FNO is the number of channels, Fourier modes and Fourier layers. In this work, the FNO was constructed using two Fourier layers, four channels and 64 Fourier modes, which correspond to half of the maximum number of modes. These choices were found to lead to accurate solutions of the forward problem while retaining relatively low computational burden. 

The neural network was implemented in Python 3.12.2 and PyTorch 2.5.1. The FNO was trained using the ADAM optimiser with a learning rate of $10^{-3}$ and a batch size of two. The FNO was trained for 400 epochs with a training time of 27 minutes per epoch and total training time of 180 hours. The FNO was trained using an Nvidia Tesla P100 PCIe GPU with 16 GB of memory. The k-Wave simulations were implemented in MATLAB R2024b using 2D GPU simulations on the same GPU.

% ============================================================================
\subsection{Forward simulation} 
\begin{figure}[tp]
    \centering
    \includegraphics[width =1\textwidth]{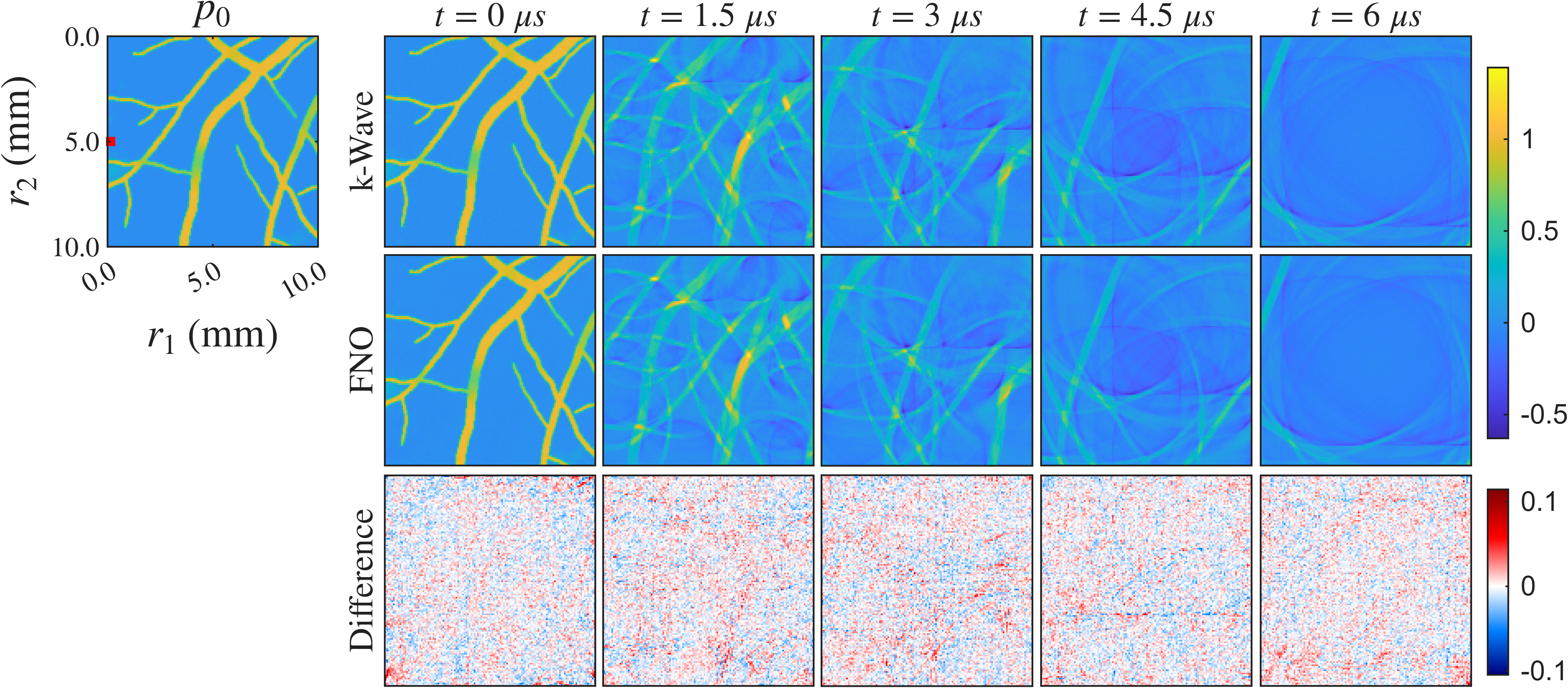}
    \caption{Simulated photoacoustic wave-field for a vessel phantom with an initial pressure distribution $p_0$ (first column) using the pseudospectral $k$-space method (k-Wave, first row) and the FNO (second row), and their difference (third row). Location used for illustration of photoacoustic data in Figure \ref{fig:timePlot} is indicated with a red dot in the first column.}
    \label{fig:FNOforwardVessel}
\end{figure}

\begin{figure}[tbp]
    \centering
    \includegraphics[width = 1\textwidth]{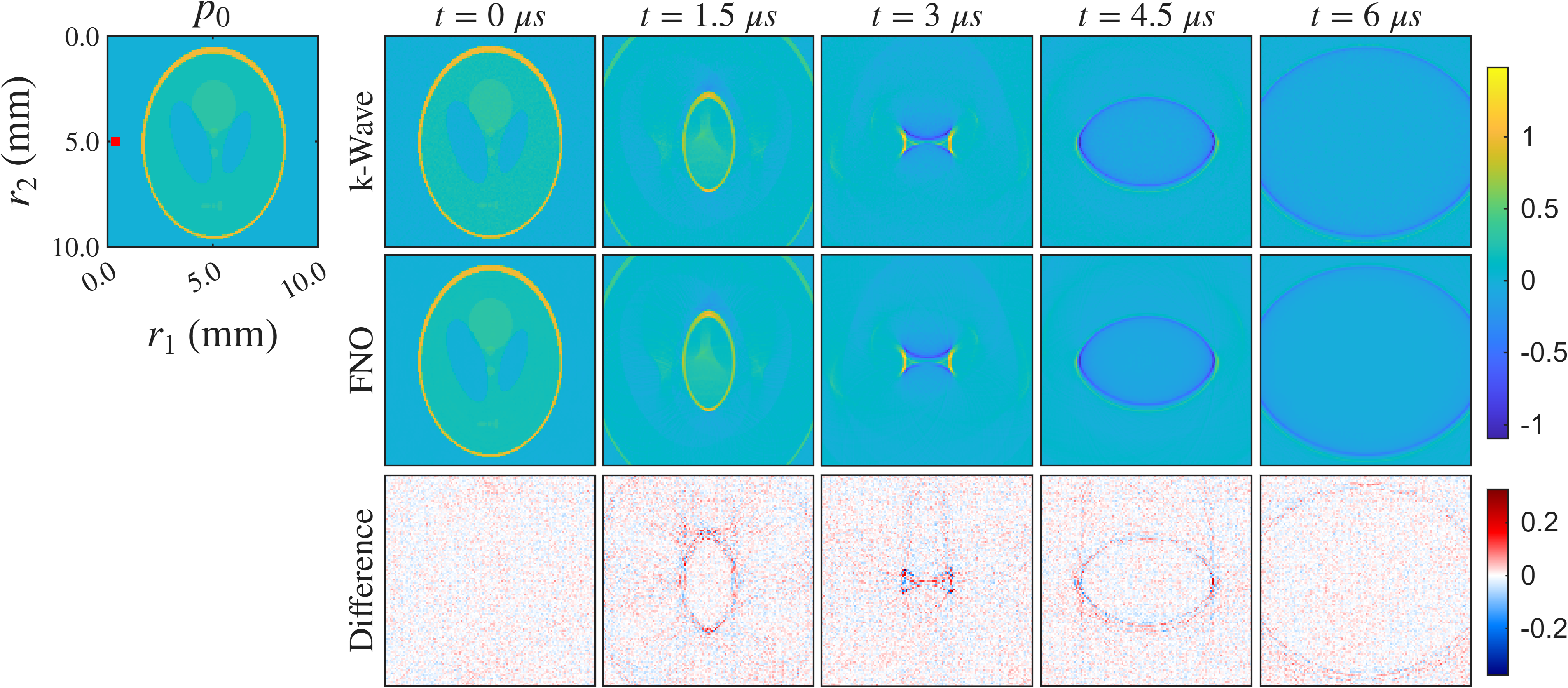}
    \caption{Simulated photoacoustic wave-field for the Shepp-Logan phantom with an initial pressure distribution $p_0$ (first column) using the pseudospectral $k$-space method (k-Wave, first row) and the FNO (second row), and their difference (third row). Location used for illustration of photoacoustic data in Figure \ref{fig:timePlot} is indicated with a red dot in the first column.}
    \label{fig:FNOforwardShepp}
\end{figure}

\begin{figure}[tbp]
    \centering
    \includegraphics[scale=0.45]{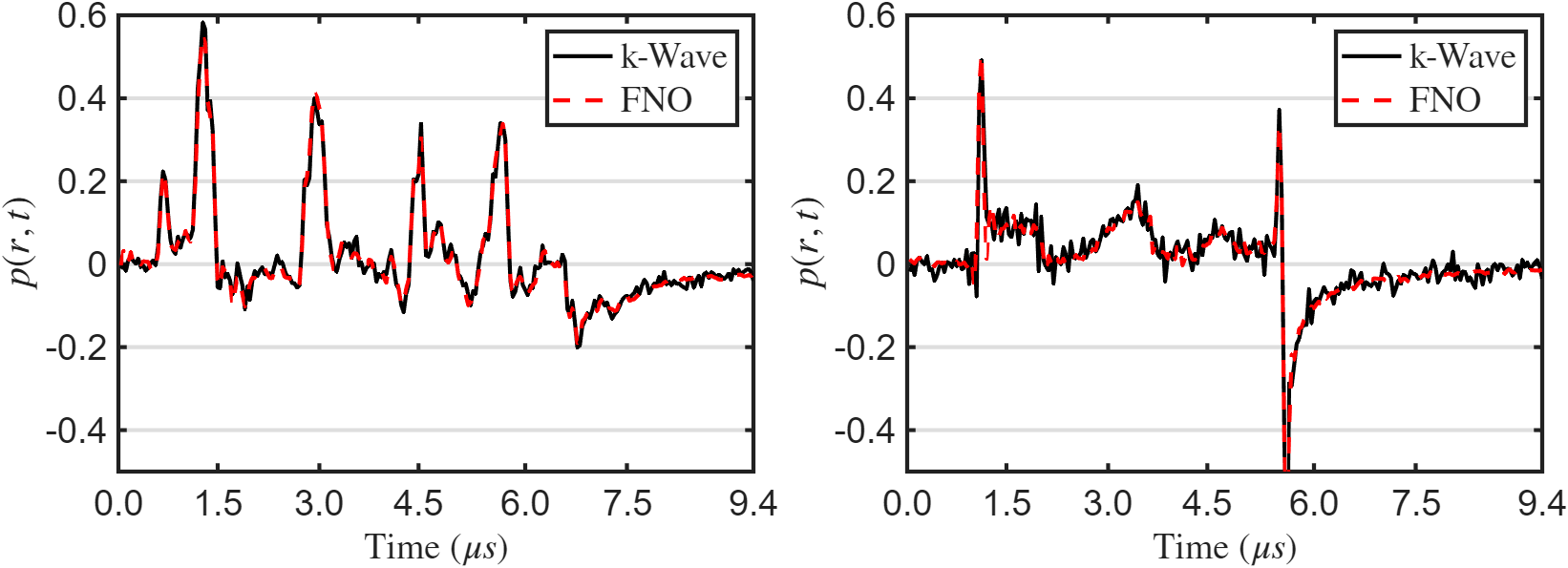}
    \caption{Photoacoustic data simulated using the pseudospectral $k$-space method (k-Wave, black line) and the FNO (red dashed line)  for the vessel (left image) and Shepp-Logan (right image) on a boundary pixel illustrated in Figures \ref{fig:FNOforwardVessel} and \ref{fig:FNOforwardShepp}. }
    \label{fig:timePlot}
\end{figure}

Simulated photoacoustic wave-field for a vessel phantom (from the testing data-set) and the Shepp-Logan phantom  (out of the training distribution data) using the pseudospectral $k$-space method and the FNO are shown in Figures \ref{fig:FNOforwardVessel} and \ref{fig:FNOforwardShepp}, respectively. Additionally, photoacoustic data on one pixel in the domain boundary (simulating measurement data at that point) are shown in Figure \ref{fig:timePlot}. 

As it can be seen from Figures \ref{fig:FNOforwardVessel} and \ref{fig:timePlot}, the FNO can approximate the photoacoustic wave-field with a similar accuracy as the pseudospectral $k$-space method. The differences can mostly be observed in the background amplitudes as small noise-like artefacts. Looking at Figures \ref{fig:FNOforwardShepp} and \ref{fig:timePlot} shows that the FNO is able to generalise well outside the training data. However, slightly larger errors compared to the vessel phantom can be seen in the high amplitude wavefronts propagating through the domain. 

Accuracy of the simulated photoacoustic wave-field using the FNO was evaluated by computing relative differences 
\begin{equation}
    \textrm{RD} = 100 \%\frac{\Vert p_t^{SIM} - p_t^{EST}\Vert}{\Vert p_t^{SIM}\Vert},
\end{equation}
where $p_t^{SIM}$ is photoacoustic wave-field simulated using the pseudospectral $k$-space method and $p_t^{EST}$ is the photoacoustic wave-field simulated using the FNO. The mean and standard deviation of the relative differences for the vessel phantom testing data was (6.16 ± 2.11) $\%$ and the relative difference of the Shepp-Logan phantom was 20.6 $\%$.
Computation times using the pseudospectral $k$-space method and the FNO were 0.44 seconds and 0.057 seconds, respectively.

% ============================================================================

\subsection{Inverse problem} \label{IP}
In solving the inverse problem, the spatial and temporal inversion discretisation described in Table \ref{SimParameters} were used. The photoacoustic datasets simulated using the pseudospectral $k$-space method in the vessel phantom and in the Shepp-Logan phantom were studied in the three different sensor geometries (the full view and two limited view geometries). The MAP estimates were solved by minimising \eqref{eq:MAP2} using the BFGS method. In solving the inverse problem, the same FNO was used in all sensor geometries. The results were compared to a reference method, where the MAP estimates were solved by minimising \eqref{eq:MAP1} using the BFGS method. In the reference method, the forward operator was implemented in a matrix-free manner utilising k-Wave toolbox in MATLAB for simulating the photoacoustic wave propagation \cite{Tick2018, Tick2016}.

The noise was modelled as Gaussian distributed zero mean noise, where the standard deviation was set to 1$\%$ of the maximum value of the photoacoustic data simulated with the vessel phantom. Same noise statistics were used for the Shepp-Logan phantom. The prior model was the Gaussian Ornstein-Uhlenbeck prior, that is defined by the covariance function \cite{Rasmussen2006}
\begin{equation}
    \Gamma_{p_0,ij} = \sigma_{p_0}^2 \exp\left\{-\frac{\Vert r_i - r_j \Vert}{l}\right\},
\end{equation}
where $\sigma_{p_0}$ is the standard deviation, $r_{i,j}$ are the locations of the discretisation points, and $l$ is the characteristic length scale controlling spatial correlation. In this work, the expected value, the standard deviation, and characteristic length were set as $\eta_{p_0} = 0.5$, $\sigma_{p_0} = 0.25$, and $l = 1$ mm, respectively. 

The initial guess for all BFGS iterations was selected to be the expected value of the prior distribution. The minimisation problem was iterated until the difference between the objective function values for five subsequent iterations was below threshold value, which was sufficient for the problem to convergence in all cases. 

Accuracy of the MAP estimates was evaluated by computing relative errors 
\begin{equation}
    \mathrm{RE} = 100 \%\frac{\Vert p_0^{SIM} - p_0^{EST}\Vert}{\Vert p_0^{SIM} \Vert},
\end{equation}
where $p_0^{SIM}$ is the true initial pressure and $p_0^{EST}$ is the estimated initial pressure.

\begin{figure}[tbp]
    \centering
    \includegraphics[width=1\textwidth]{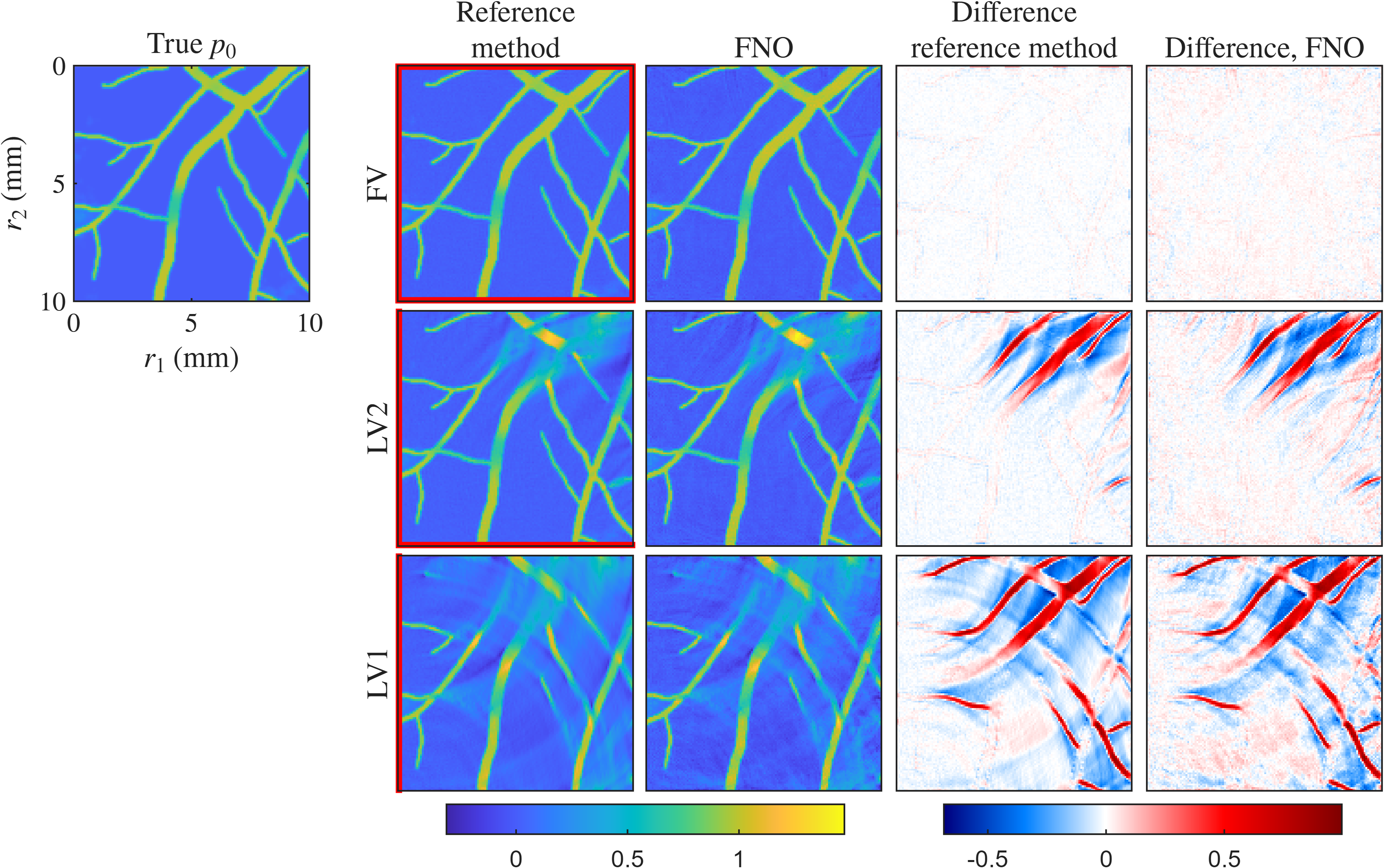}
    \caption{The MAP estimates computed from a vessel phantom data in different sensor geometries: full-view (FV, first row), two-side limited view (LV2, second row) and one-side limited view (LV1, third row). Columns from left to right: the true initial pressure distribution $p_0$ (first column), MAP estimates calculated using the reference method (second column), MAP estimates calculated using the FNO and automatic differentation (third column), the difference between the true and the estimated initial pressure for the reference method (fourth column), and the difference between true and estimated initial pressure for the FNO approach (fifth column). Sensor locations are marked with the red line in the second column.}
    \label{fig:InverseVesselImage}
\end{figure}

The MAP estimates for one sample of the vessel phantom calculated in three sensor geometries are shown in Figure \ref{fig:InverseVesselImage}, and the relative errors for the testing vessel dataset are shown in Table \ref{reInverse}. As it can be seen, the MAP computed with the FNO and the reference method appear visually similar in all sensor geometries, including the blurring of structures characteristic to limited view sensor geometries. Furthermore, the relative errors of the MAP estimates are of the similar magnitude for the FNO and the reference method in all sensor geometries. 

\begin{table}[tbp]
\centering
\caption{Mean $\pm$ standard deviation of the relative errors (RE) of the MAP estimates the vessel testing dataset and the Shepp-Logan phantom calculated from photoacoustic data in a full-view (FV), two-side (LV2) and one-side (LV1) sensor geometries.}
\begin{tabular}{lccccccccc}
\hline
    &  \multicolumn{3}{c}{RE (\%), vessel} &  &  \multicolumn{3}{c}{RE (\%), Shepp-Logan} \\
    &  Reference method             &  & FNO             &  &  Reference method            &            & FNO          \\ \hline
FV  & 4.58 $\pm$ 1.03 & & 4.63 $\pm$ 1.54 & & 15.5 & & 16.5    \\
LV2 & 13.4 $\pm$ 6.08 & & 13.8 $\pm$ 6.33 & & 20.5 & & 21.5    \\
LV1 & 33.4 $\pm$ 12.4 & & 34.1 $\pm$ 12.5 & & 50.1 & & 51.3    \\ \hline
\end{tabular}
\label{reInverse}
\end{table}

\begin{figure}[h]
    \centering
    \includegraphics[width=1\textwidth]{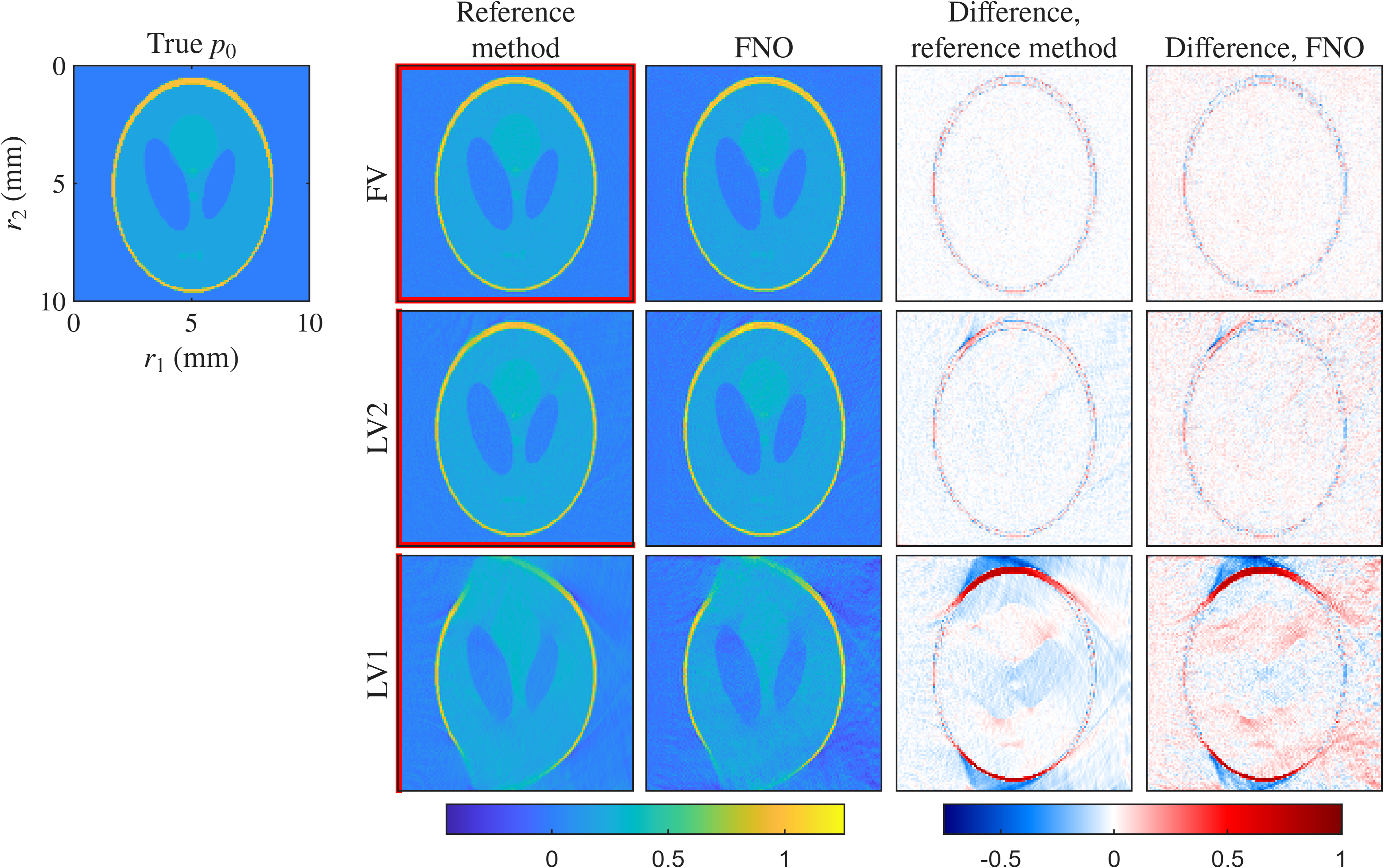}
    \caption{The MAP estimates computed from the Shepp-Logan phantom data in different sensor geometries: full-view (FV, first row), two-side limited view (LV2, second row) and one-side limited view (LV1, third row). Columns from left to right: the true initial pressure distribution $p_0$ (first column), MAP estimates calculated using the reference method (second column), MAP estimates calculated using the FNO and automatic differentation (third column), the difference between the true and the estimated initial pressure for the reference method (fourth column), and the difference between true and estimated initial pressure for the FNO approach (fifth column). Sensor locations are marked with the red line in the second column.}
    \label{fig:InverseSheppImage}
\end{figure}

The MAP estimates calculated using data simulated with the Shepp-Logan phantom in the three sensor geometries are shown in Figure \ref{fig:InverseSheppImage}, and the corresponding relative errors are given in Table \ref{reInverse}. As it can be seen, only small differences can be observed in the MAP estimates calculated in the full-view and two-side sensor geometries, whereas slight increase in the error can be seen with the one-side sensor geometry. Furthermore, as in the case of the testing vessel dataset, the relative error are of the similar magnitude between the FNO and the reference method solutions.

\subsection{Convergence and computation time} 
\begin{figure}[h]
    \centering
    \includegraphics[width=1\textwidth]{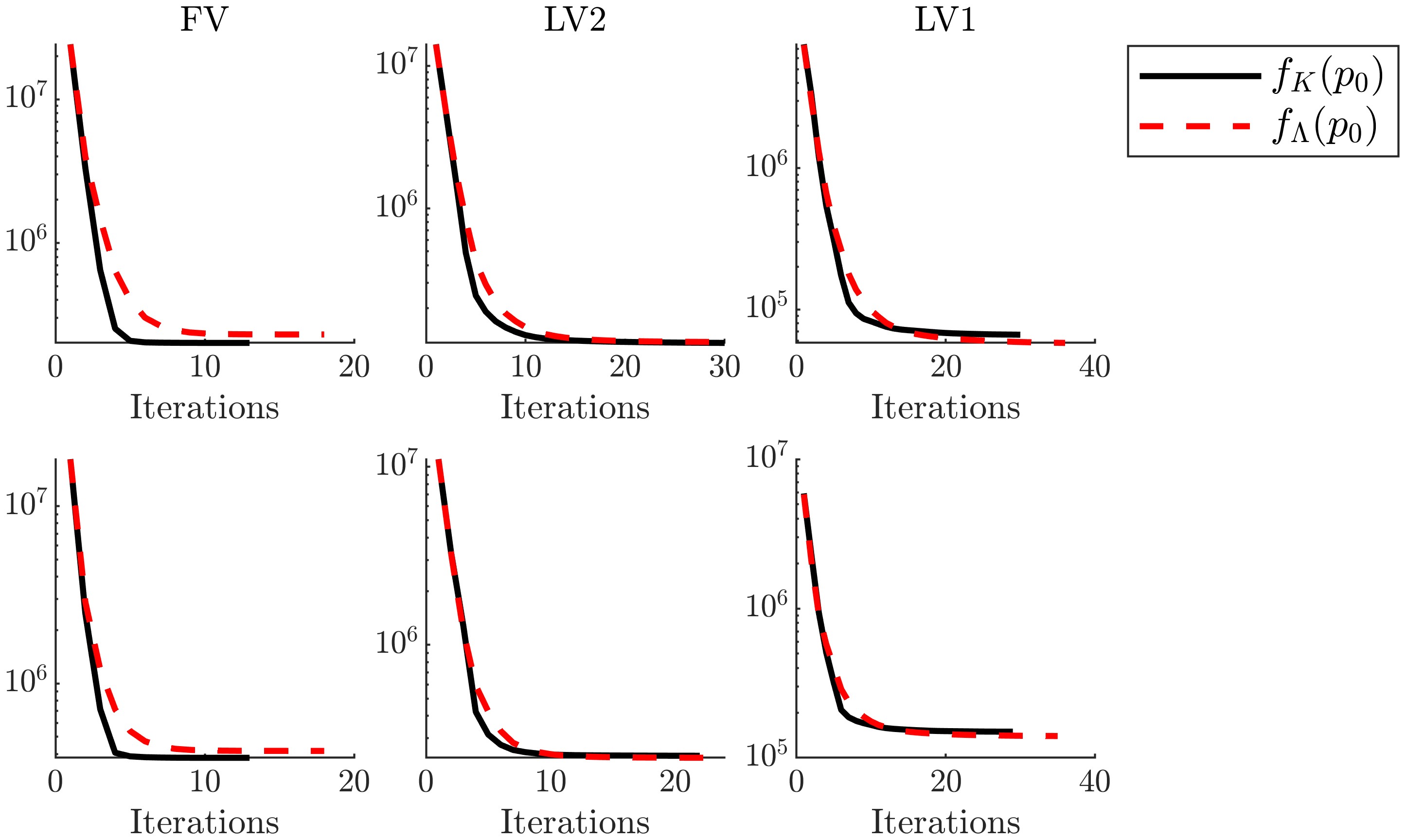}
    \caption{Objective function values against iteration during the BFGS algorithm when the reference method (black line) and the FNO (red dashed line) are used for the vessel data (first row) and for the Shepp-Logan (second row) using three different sensor geometries: full-view (FV, first column), two-sided limited view (LV2, second column), and one-sided limited view (LV1, third column).}
    \label{fig:objFunct}
\end{figure}

The objective function values during the iterations of the BFGS algorithm for the results presented in Figures \ref{fig:InverseVesselImage} and \ref{fig:InverseSheppImage} are shown in Figure \ref{fig:objFunct}. As it can be observed, all of the minimisation problems converged in approximately the same number of iterations, regardless of the forward operator or the sensor geometry. However, in some cases, the FNO did not reach as low value of the objective function as the reference method, which could be due small modelling errors of the FNO. 

Computation times of the gradient using the reference method \eqref{eq:Gradient1} and the FNO \eqref{eq:Gradient2} are shown in Table \ref{IterationTimes}. As it can be observed, the computation times for the gradients using the FNO decrease with smaller number of sensors even thought the same neural network is used in all of them. On the other hand, computing the gradients using the reference method is slower and remains constant despite changes in sensor geometry. 

\begin{table}[tbp]
\centering
\caption{Computation time of the gradients using the reference method and the FNO for the three sensor geometries.}
\begin{tabular}{lclcclcclc}
\hline
    &  &  & \multicolumn{7}{c}{Computation time (s)} \\
    & & & FV  & & & LV2  & & & LV1  \\ \hline
Reference method & & &  0.47 & & &  0.47  & & &  0.47  \\
FNO & & &  0.0069  & & &  0.0063 & & &  0.0062  \\ \hline
\end{tabular}
\label{IterationTimes}
\end{table}

\section{Discussion and conclusions} \label{5}
In this work, an approach to the inverse problem of PAT using a FNO-based learned forward operator was presented. The inverse problem was solved using the BFGS method where the gradient was approximated using automatic differentiation. The  MAP estimates calculated utilising the learned forward operator were compared to the reference method where the forward operator was approximated using the pseudospectral $k$-space method implemented with the k-Wave toolbox in MATLAB. The simulations demonstrated that the FNO was able to approximate the solution of the wave equation with a similar accuracy as the pseudospectral $k$-space method. Furthermore, solving the inverse problem of PAT using the FNO and automatic differentiation provided equally good estimates of the initial pressure as the conventional reference approach. Moreover, computation times of the gradients and simulation of the ultrasound data using the FNO were significantly shorter compared to the reference method.

In this work, the FNO was trained to estimate the solution of a linear model. The methodology could, however, be utilised in other, possibly non-linear, forward models and inverse problems as long as a sufficiently accurate neural network models can be trained. Automatic differentiation further enables a computationally efficient approach for computing derivatives such as Jacobian matrices, which would enable the use of other optimisation methods as well, for example, Gauss-Newton method. The approach is also flexible in terms of the chosen neural network architecture as long as a sufficiently accurate model of the forward operator can be trained. The advantages of the use of the FNO in PAT include its ability to simulate ultrasound propagation across the entire domain quickly compared to the pseudospectral $k$-space method. Additionally, once trained, the FNO can be used in the inverse problem with arbitrary sensor geometries and finitely sized sensors. 

In this work, the inverse problem of PAT was studied using 2D simulations. Extending the proposed method to 3D using the FNO can be challenging due to increasing memory requirements. To alleviate this, other computationally more efficient neural network architectures could be used such as ones that only model the ultrasound data for a chosen sensor points. Furthermore, the proposed approach was tested using simulated data and simple sensor geometries. To move towards more realistic applications, the approach should be tested using experimental data and sensor models taking into account factors such as directivity, frequency response and finite size.

Overall, the results show that the FNO can be used to approximate the propagation of photoacoustic wave-field with a similar accuracy as conventional numerical methods. Furthermore, it can be utilised in the inverse problem of PAT with automatic differentiation to estimate the initial pressure both in full view and limited view sensor geometries. It was also demonstrated that the methodology is able to generalise well to a target outside the training data. 

\section*{Acknowledgments}
This project received funding from the Finnish Ministry of Education and Culture’s Pilot for Doctoral Programmes (Pilot project Mathematics of Sensing, Imaging and Modelling),  the European Research Council (ERC) under the European Union’s Horizon 2020 research and innovation programme (grant agreement No 101001417 - QUANTOM), the Research Council of Finland (Centre of Excellence in Inverse Modelling and Imaging grant (353086, 353093), Flagship of Advanced Mathematics for Sensing Imaging and Modelling grant (358944, 359186), and Flagship Program Photonics Research and Innovation grant 320166), Orion Research Foundation, and Academy Research Fellow (AI-SOL) grant 338408. The authors wish to acknowledge CSC - IT Center for Science, Finland, for computational resources.

\bibliographystyle{plain}  
\bibliography{references}

@article{Allman2018,
    author = {Allman, D. and Reiter, A. and Bell, M. A. L.},
    title = {Photoacoustic Source Detection and Reflection Artifact Removal Enabled by Deep Learning}, 
    journal = {IEEE Transactions on Medical Imaging}, 
    publisher = {},
    volume = {37},
    number = {6},
    pages = {1464-1477},
    year = {2018},
    doi = {10.1109/TMI.2018.2829662},
}

@article{Antholzer2019,
    author = {Antholzer, S. and Haltmeier, M. and Schwab, J.},
    title = {Deep Learning for Photoacoustic Tomography from Sparse Data},
    journal = {Inverse Problems in Science and Engineering},
    volume = {27},
    number = {7},
    pages = {987–1005},
    year = {2019},
    doi = {10.1080/17415977.2018.1518444}
}

@article{Arridge2016,
    author = {Arridge, S. and Beard, P. and Betcke, M. and Cox, B. and Huynh, N. and Lucka, F. and Ogunlade, O. and Zhang, E.},
    title = {Accelerated high-resolution photoacoustic tomography via compressed sensing},
    journal = {Physics in Medicine and Biology},
    publisher = {IOP Publishing},
    volume = {61},
    number = {24},
    pages = {8908-8940},
    year = {2016},
    doi = {10.1088/1361-6560/61/24/8908}
}

@article{Arridge2019, 
    author = {Arridge, S. and Maass, P. and Öktem, O. and Schönlieb, C-B.}, 
    title = {Solving inverse problems using data-driven models}, 
    journal = {Acta Numerica}, 
    volume = {28}, 
    pages = {1–174},
    year = {2019}, 
    doi = {10.1017/S0962492919000059}
}

@article{Beard2011,
    author = {Beard, P.},
    title = {Biomedical photoacoustic imaging},
    journal = {Interface Focus},
    volume = {1},
    pages = {602-631},
    year = {2011},
    doi = {10.1098/rsfs.2011.0028}
}

@article{Budai2013,
    author = {Budai, A. and Bock, R. and Maier, A. and Hornegger, J. and Michelson, G.},
    title = {Robust vessel segmentation in fundus images},
    journal = {International Journal of Biomedical Imaging},
    volume = {2013},
    number = {1},
    pages = {154860},
    year = {2013},
    doi = {10.1155/2013/154860}
}

@article{Burgholzer2007,
    author = {Burgholzer, P. and Matt, G. J. and Haltmeier, M. and Paltauf, G.},
    title = {Exact and approximative imaging methods for photoacoustic tomography using an arbitrary detection surface},
    journal = {Phys. Rev. E},
    publisher = {American Physical Society},
    volume = {75},
    issue = {4},
    pages = {046706},
    numpages = {10},
    year = {2007},
    doi = {10.1103/PhysRevE.75.046706}
}

@article{Cox2004,
    author = {Cox, B. T. and Beard, P. C.},
    title = {Fast calculation of pulsed photoacoustic fields in fluids using k-space methods},
    journal = {J. Acoust. Soc. Am.},
    publisher = {},
    volume = {117},
    number = {6},
    pages = {3616–3627},
    numpages = {},
    year = {2004},
    doi = {}
}

@article{Cox2007,
    author = {Cox, B. T. and Arridge, S. R. and Beard, P. C.},
    title = {Photoacoustic tomography with a limited-aperture planar sensor and a reverberant cavity},
    journal = {Inverse Problem},
    publisher = {American Physical Society},
    volume = {23},
    issue = {},
    pages = {S95-S112},
    numpages = {},
    year = {2007},
    doi = {10.1088/0266-5611/23/6/S08}
}

@article{Cox2007b,
    author = {Cox, B. T. and Kara S. and Arridge, S. R. and Beard, P. C.},
    title = {k-space propagation models for acoustically heterogeneous media: application to biomedical photoacoustics},
    journal = {J. Acoust. Soc. Am.},
    publisher = {},
    volume = {121},
    number = {6},
    pages = {3453-64},
    year = {2007},
    doi = {10.1121/1.2717409}
}

@article{Dean-Ben2012,
  author={Deán-Ben, X. L. and Buehler, A. and Ntziachristos, V. and Razansky, D.},
  journal={IEEE Transactions on Medical Imaging}, 
  title={Accurate Model-Based Reconstruction Algorithm for Three-Dimensional Optoacoustic Tomography}, 
  year={2012},
  volume={31},
  number={10},
  pages={1922-1928},
  doi={10.1109/TMI.2012.2208471}}

@article{Finch2004,
  author = {Finch, D. and  Patch, S. K. and Rakesh},
  journal = {SIAM Journal on Mathematical Analysis}, 
  title = {Determining a Function from its Mean Values Over a Family of Spheres}, 
  year = {2004},
  volume = {35},
  number = {},
  pages = {1213–1240},
  doi = {10.1109/TMI.2012.2208471}}

@article{Frikel2018,
  author = {Frikel, J. and Haltmeier, M.},
  journal = {Inverse Problems}, 
  title = {Efficient regularization with wavelet sparsity constraints in photoacoustic tomography}, 
  year = {2018},
  volume = {34},
  number = {2},
  pages = {024006},
  doi = {10.1088/1361-6420/aaa0ac}}

@article{Grohl2021,
   author = {Gröhl, J. and Schellenberg, M. and Dreher, K. and Maier-Hein, L.},
   title = {Deep learning for biomedical photoacoustic imaging: A review},
   journal = {Photoacoustics},
   publisher = {Elsevier BV},
   volume = {22},
   ISSN = {2213-5979},
   pages = {100241},
   year = {2021},
   doi = {10.1016/j.pacs.2021.100241}
}

@article{Guan2023,
    author = {Guan, S. and Hsu, K.-T. and Chitnis, P.},
    title = {Fourier neural operator network for fast photoacoustic wave simulations},
    journal = {Algorithms},
    volume = {16}, 
    number = {2},
    pages = {124},
    year = {2023},
    doi = {10.3390/a16020124}
}

@article{Guan2020,
  author = {Guan, S. and Khan, A. A. and Sikdar, S. and Chitnis, P. V.},
  title = {Fully Dense \textsc{UN}et for 2-\textsc{D} Sparse Photoacoustic Tomography Artifact Removal}, 
  journal = {IEEE Journal of Biomedical and Health Informatics}, 
  volume = {24},
  number = {2},
  pages = {568-576},
  year = {2020},
  doi = {10.1109/JBHI.2019.2912935}
}

@article{Hauptmann2020,
    author = {Hauptmann, A. and Cox, B.},
    title = {Deep learning in photoacoustic tomography: \textsc{C}urrent approaches and future directions},
    journal = {Journal of Biomedical Optics},
    publisher = {SPIE-Intl Soc Optical Eng},
    volume = {25},
    number = {11},
    pages = {112903},
    ISSN = {1083-3668},
    year = {2020},
    DOI = {10.1117/1.JBO.25.11.112903},
}

@article{Hauptmann2018,
    author = {Hauptmann, A. and Lucka, F. and Betcke, M. and Huynh, N. and Adler, J. and Cox, B. and Beard, P. and Ourselin, S. and Arridge, S.},
    title = {Model-Based Learning for Accelerated, Limited-View 3-\textsc{D} Photoacoustic Tomography}, 
    journal = {IEEE Transactions on Medical Imaging}, 
    volume = {37},
    number = {6},
    pages = {1382-1393},
    year = {2018},
    doi = {10.1109/TMI.2018.2820382}
}

@article{Hristova2008,
    author = {Hristova, Y. and Kuchment, P. and Nguyen, L.},
    title = {Reconstruction and time reversal in thermoacoustic tomography in acoustically homogeneous and inhomogeneous media},
    journal = {Inverse Problems},
    volume = {24},
    number = {5},
    pages = {055006},
    year = {2008},
    doi = {10.1088/0266-5611/24/5/055006}
}

@article{Hsu2023,
    author = {Hsu, K.-T. and Guan, S. and Chitnis, P. V. },
    title = {Fast iterative reconstruction for photoacoustic tomography using learned physical model: Theoretical validation},
    journal = {Photoacoustics},
    volume = {29},
    pages = {100452},
    issn = {2213-5979},
    year = {2023},
    doi = {10.1016/j.pacs.2023.100452}
}

@article{Haltmeier2005,
    author = {Haltmeier, M. and Schuster, T. and Scherzer, O. },
    title = {Filtered backprojection for thermoacoustic computed tomography in spherical geometry},
    journal = {Mathematical Methods in the Applied Sciences},
    volume = {28},
    pages = {1919-1937},
    issn = {},
    year = {2005},
    doi = {10.1002/mma.648}
}

@article{Kaipio2005,
    author = {Kaipio, J. P. and Somersalo, E.},
    title = {Statistical and Computational Inverse
    Problems},
    journal = {Springer-Verlag},
    year = {2005},
    doi = {10.1007/b138659}
}

@article{Kultima2026,
    author = {Kultima, J. and Ramlau, R. and Sahlström, T. and Tarvainen, T.},
    title = {Fast reconstruction approaches for photoacoustic tomography with smoothing \textsc{S}obolev/\textsc{M}atérn priors},
    pages = {},
    volume = {accepted for publication},
    number = {},
    year = {},
    journal = {Inverse Problems and Imaging},
    doi = {}
}

@article{Kunyansky2007,
    author = {Kunyansky, L. A.},
    title = {Explicit inversion formulae for the spherical mean \textsc{R}adon transform},
    pages = {373-383},
    volume = {23},
    number = {1},
    year = {2007},
    journal = {Inverse Problems},
    doi = {10.1088/0266-5611/23/1/021}
}

@article{Kuchment2008,
title = {A survey in mathematics for industry: Mathematics of thermoacoustic tomography},
author = {P. Kuchment and L. Kunyansky},
year = {2008},
volume = {19},
pages = {191-224},
journal = {European Journal of Applied Mathematics},
publisher = {Cambridge University Press},
number = {2},
}

@article{Li2021,
    author = {Li, Z. and Kovachki, N. and Azizzadenesheli, K. and Liu, B. and Bhattacharya, K. and Stuart, A. and Anandkumar, A.},
    title = {Fourier Neural Operator for Parametric Partial Differential Equations},
    journal = {International Conference on Learning Representations (ICLR)},
    year = {2021},
    doi = {}
}

@article{Lu2021,
    author = {Lu, L. and Jin, P. and Pang, G. and Zhang, Z. and G. Karniadakis, E.},
    title = {Learning nonlinear operators via \textsc{D}eep\textsc{ON}et based on the universal approximation theorem of operators},
    journal = {Nat Match Intell},
    volume = {3},
    number = {},
    pages = {218-229},
    year = {2021},
    doi = {ttps://doi.org/10.1038/s42256-021-00302-5}
}

@article{Mitsuhashi2017,
    author = {Mitsuhashi, K. and Poudel, J. and Matthews, T. P. and Garcia-Uribe, A. and Wang, L. V. and Anastasio, M. A.},
    title = {A forward-adjoint operator pair based on the elastic wave equation for use in transcranial photoacoustic computed tomography},
    journal = {SIAM Journal of Imaging Sciences},
    volume = {10},
    number = {4},
    pages = {2022–2048},
    year = {2017}, 
    doi = {10.1137/16M1107619}
}

@book{Nocedal2006,
    author = {Nocedal, J. and Wright, S.},
    title = {Numerical Optimization}, 
    publisher = {Springer}, 
    year = {2006}, 
    doi = {10.1007/978-0-387-40065-5}
}

@article{Poudel2019,
    author = {Poudel, J. and Lou, Y. and Anastasio, M. A.},
    title = {A survey of computational frameworks for solving the acoustic inverse problem in three-dimensional photoacoustic computed tomography},
    journal = {Physics in Medicine \& Biology},
    publisher = {IOP Publishing},
    volume = {64},
    number = {14},
    pages = {14TR01},
    year = {2019},
    doi = {10.1088/1361-6560/ab2017}
}

@article{Raissi2019,
    author = {Raissi, M. and Perdikaris, P. and Karniadakis, G. E. },
    title = {Physics-informed neural networks: A deep learning framework for solving forward and inverse problems involving nonlinear partial differential equations},
    journal = {Journal of Computational Physics},
    volume = {378},
    pages = {686-707},
    year = {2019},
    issn = {0021-9991},
    doi = {https://doi.org/10.1016/j.jcp.2018.10.045}
}

@article{Rasmussen2006,
    author = {C. E. Rasmussen and C. K. I. Williams},
    title = {Gaussian Processes for Machine Learning},
    journal = {MIT Press},
    year = {2006},
    month = {},
    pages = {},
    volume = {MA},
    doi = {https://doi.org/10.7551/mitpress/3206.001.0001}
}

@article{Sahlstrom2023,
    author = { Sahlstr\"{o}m, T. and Tarvainen, T. },
    title = {Utilizing variational autoencoders in the \textsc{B}ayesian inverse problem of photoacoustic tomography},
    journal = {SIAM Journal on Imaging Sciences},
    volume = {16},
    number = {1},
    pages = {89-110},
    year = {2023},
    doi = {10.1137/22M1489897}
}

@article{Sahlstrom2020,
    author = {Sahlstr\"{o}m, T. and Pulkkinen, A. and Tick, J. and Leskinen, J. and Tarvainen, T.},
    title = {Modeling of Errors Due to Uncertainties in Ultrasound Sensor Locations in Photoacoustic Tomography},
    journal = {IEEE Transactions on Medical Imaging},
    volume = {39},
    number = {6},
    pages = {2140-2150},
    year = {2020},
    doi = {10.1109/TMI.2020.2966297}
}

@article{Tick2016,
    author = {Tick, J. and Pulkkinen, A. and Tarvainen, T.},
    title = {Image reconstruction with uncertainty quantification in photoacoustic tomography},
    journal = {J. Acoust. Soc. Am.},
    volume = {139},
    number= {4},
    pages = {1951-1961},
    year = {2016},
    doi = {10.1121/1.4945990}
}

@article{Tick2018,
    author = {Tick, J. and Pulkkinen, A. and Lucka, F. and Ellwood, R. and Cox, B. T. and Kaipio, J. P. and Arridge, S. R. and Tarvainen, T.},
    title = {Three dimensional photoacoustic tomography in Bayesian framework},
    journal = {J. Acoust. Soc. Am.},
    publisher = {},
    volume = {144},
    number = {4},
    pages = {2061–2071},
    year = {2018},
    doi = {10.1121/1.5057109}
}

@article{Treeby2010Kwave,
    author = {Treeby, B. and Cox, B. T.},
    title = {k-\textsc{W}ave: \textsc{MATLAB} toolbox for the simulation and reconstruction of photoacoustic wave fields},
    journal = {Journal of Biomedical Optics},
    pages = {021314},
    volume = {15},
    number = {2},
    year = {2010},
    doi = {10.1117/1.3360308}
}

@article{Treeby2010,
    author = {Treeby, B. E. and Zhang, E. Z. and Cox, B. T.},
    title = {Photoacoustic tomography in absorbing acoustic media using time reversal},
    journal = {Inverse Problems}, 
    publisher = {},
    volume = {26},
    number = {11},
    pages = {115003},
    year = {2010},
    doi = {10.1088/0266-5611/26/11/115003}
}

@inproceedings{Waibel2018,
    author = {Waibel, D. and Gr{\"o}hl, J. and Isensee, F. and Kirchner, T. and Maier-Hein, K. and Maier-Hein, L.},
    title = {Reconstruction of initial pressure from limited view photoacoustic images using deep learning},
    booktitle = {Photons Plus Ultrasound: Imaging and Sensing 2018},
    organization = {International Society for Optics and Photonics},
    publisher = {SPIE},
    volume = {10494},
    pages = {104942S},
    year = {2018},
    doi = {10.1117/12.2288353}
}

@article{Wang2013,
    author = {Wang, K. and Huang, C. and Kao, Y.-J. and Chou, C.-Y. and Oraevsky, A. A and Anastasio, M. A},
    title = {Accelerating image reconstruction in three-dimensional optoacoustic tomography on graphics processing units},
    journal = {Physics in Medicine},
    publisher = {},
    volume = {40},
    number = {2},
    pages = {023301},
    year = {2013},
    doi = {10.1118/1.4774361}
}

@article{Wang2012,
    author = {Wang, K. and Su, R. and Oraevsky, A. A. and Anastasio, M. A.},
    title = {Investigation of iterative image reconstruction in three-dimensional optoacoustic tomography},
    journal = {Physics in Medicine \& Biology},
    publisher = {IOP Publishing},
    volume = {57},
    number = {17},
    pages = {5399-5423},
    year = {2012},
    doi = {10.1088/0031-9155/57/17/5399}
}

@article{Wang2016,
    author = {Wang, L. and Yao, J.},
    title = {A practical guide to photoacoustic tomography in the life sciences},
    journal = {Nature Methods},
    pages = {627-638},
    volume = {13},
    number = {8},
    year = {2016},
    doi = {10.1038/nmeth.3925}
}

@article{Xu2005,
    author = {Xu, M. and Wang, L. V.},
    title = {Universal back-projection algorithm for photoacoustic computed tomography},
    journal = {Phys. Rev. E},
    publisher = {American Physical Society},
    volume = {71},
    issue = {1},
    pages = {016706},
    numpages = {7},
    year = {2005},
    doi = {10.1103/PhysRevE.71.016706}
}

@article{Xu2003,
    author = {Xu, M. and Xu, Y. and Wang, L. V.},
    title = {Time-domain reconstruction algorithms and numerical simulations for thermoacoustic tomography in various geometries},
    journal = {IEEE Transactions on Biomedical Engineering},
    publisher = {},
    volume = {50},
    number = {9},
    pages = {1086-1099},
    year = {2003},
    doi = {10.1109/TBME.2003.816081}
}

@article{Xu2004,
    author = {Xu, Y. and Wang, L. V.},
    title = {Time Reversal and Its Application to Tomography with Diffracting Sources},
    journal = {Phys. Rev. Lett.},
    publisher = {American Physical Society},
    volume = {92},
    issue = {3},
    pages = {033902},
    numpages = {4},
    year = {2004},
    doi = {10.1103/PhysRevLett.92.033902}
}

@article{Yang2020,
    author = {Yang, C.  and Lan, H. and Gao, F. and Gao, F.},
    title = {Review of deep learning for photoacoustic imaging},
    journal = {Photoacoustics},
    year = {2020},
    volume = {21}, 
    pages = {100215},
    doi = {10.1016/j.pacs.2020.100215}
}

\end{document}